\documentclass[cernpreprint, atlasdraft=false, UKenglish, texmf, orcidlogo]{atlasdoc}
 
\usepackage[subfigure]{atlaspackage}
\usepackage{atlasbiblatex}
 
\usepackage{atlasphysics}

\graphicspath{{logos/}{figures/}}
 
\usepackage{ANA-HIGG-2018-44-PAPER-defs}
\usepackage{siunitx}
\usepackage{multirow}

\AtlasTitle{Search for the $Z\gamma$ decay mode of new high-mass resonances in $pp$ collisions at $\sqrt{s}=13$~TeV with the ATLAS detector}

\AtlasAbstract{
This letter presents a search for narrow, high-mass resonances in the $Z\gamma$ final state with the $Z$ boson decaying into a pair of electrons or muons. The $\sqrt{s}=13$~TeV $pp$ collision data were recorded by the ATLAS detector at the CERN Large Hadron Collider and have an integrated luminosity of 140~fb$^{-1}$. The data are found to be in agreement with the Standard Model background expectation. Upper limits are set on the resonance production cross section times the decay branching ratio into $Z\gamma$. For spin-0 resonances produced via gluon--gluon fusion, the observed limits at 95\% confidence level vary between 65.5~fb and 0.6~fb, while for spin-2 resonances produced via gluon--gluon fusion (or quark--antiquark initial states) limits vary between 77.4 (76.1)~fb and 0.6 (0.5)~fb, for the mass range from 220~GeV to 3400~GeV.
}
 
\author{The ATLAS Collaboration}
 
\AtlasRefCode{HIGG-2018-44}
 
\PreprintIdNumber{CERN-EP-2023-153}

\AtlasJournal{Phys.\ Lett.\ B.}
\AtlasJournalRef{Phys. Lett. B 848 (2024) 138394}
\AtlasDOI{10.1016/j.physletb.2023.138394}

\hypersetup{pdftitle={ATLAS document},pdfauthor={The ATLAS Collaboration}}
 
\begin{document}
 
\maketitle

\section{Introduction}
\label{sec:intro}

Theories beyond the Standard Model (BSM theories) predict the existence of new heavy bosons ($X$) as additional gauge fields or by expanding the Higgs sector~\cite{TheoryTC,LittleHiggs,Pappadopulo:2014qza}. High-energy proton--proton ($pp$) collisions at the CERN Large Hadron Collider (LHC) could produce high-mass bosons with masses up to several \TeV, allowing a wide range of BSM scenarios to be tested. This letter targets narrow spin-0 and spin-2 high-mass resonances that decay into the $Z\gamma$ final state, with the $Z$ boson decaying into an electron or muon pair. The $Z(\to\ell\ell)\gamma$ final state, where $\ell=e\text{ or }\mu$, can be fully reconstructed with high efficiency and the invariant mass can be measured with good resolution. In addition, lepton and photon signatures lead to relatively small backgrounds from general hadronic final-state decays.
 
The ATLAS~\cite{ATLASDET-2008, ATLAS-TDR-2010-19} and CMS~\cite{CMS-CMS-00-001,} collaborations have searched for heavy $Z\gamma$ resonances. The ATLAS Collaboration has used 36.1~\invfb\ of 13~\TeV\ $pp$ collision data to search for spin-0 and spin-2 high-mass resonances via the $Z(\to\ell\ell)\gamma$ final state. For the spin-0 resonance, the observed upper limits set on the production cross section times branching ratio, $\sigxbr$, vary between 88~fb and 2.8~fb at 95\% confidence level in the mass range from 0.25~\TeV\ to 2.4~\TeV~\cite{HIGG-2016-14}. In addition to the search in the leptonic final state, a search for hadronic $Z$ boson decays~\cite{HDBS-2019-10} was also published by the ATLAS Collaboration using a dataset with an integrated luminosity of 139 fb$^{-1}$. Upper limits from 10~fb to 0.05~fb were set in the mass range from 1.0~\TeV\ to 6.8~\TeV. A similar analysis was carried out by the CMS Collaboration with an integrated luminosity of 35.9 fb$^{-1}$. The $Z$ boson was studied in both the leptonic and hadronic decay modes. The results from these channels were combined, and for narrow spin-0 resonances with masses between 0.35~\TeV\ and 4.0~\TeV, the upper limits ranged from 50~fb to 0.3~fb~\cite{CMS-EXO-17-005}. In all these searches, the data were found to agree with the Standard Model (SM) background expectation.
 
An improved search for high-mass $Z\gamma$ resonances is presented in this letter. For bosons with masses of around a \TeV\ or above, a highly boosted $Z$ boson is produced, where the leptons from the $Z$ boson decay are highly collimated. Due to this boost, the energy deposits from the two electrons in $Z\to ee$ decays are very close together,
with an angular separation of around 0.2~rad for higher signal masses, which affects the reconstruction of individual electrons. Consequently, the use of conventional electron identification requirements causes a significant loss of signal efficiency (around 20\% for a resonance mass of 3400~\GeV). In addition, due to the closeness of the two electrons, about 20\% of the electrons are not reconstructed properly but instead classified as photons. Such challenges are addressed by developing a customized electron identification algorithm based on multivariate analysis (MVA) techniques.
The main background in this analysis is non-resonant production of a $Z$ boson with a photon. In addition, a smaller contribution comes from the production of a $Z$ boson together with a hadronic jet, when the jet is incorrectly identified as a photon. The search is based on 13~\TeV\ $pp$ collision data recorded by the ATLAS detector at the LHC from 2015 to 2018, with a total integrated luminosity of 140~\invfb. This is much larger than the dataset used in the previous ATLAS search in the $Z(\to\ell\ell)\gamma$ decay channel~\cite{HIGG-2016-14} and together with better electron identification performance, the search range is widened to cover masses from  220~\GeV\ to 3400~\GeV.


\section{ATLAS detector and data sample}
\label{sec:detectorAndData}

The ATLAS detector~\cite{ATLASDET-2008, ATLAS-TDR-2010-19} at the LHC is a multipurpose particle detector with a front-to-back symmetric cylindrical geometry and solid angle coverage close to $4\pi$.\footnote{ATLAS uses a right-handed coordinate system with its origin at the nominal interaction point (IP)
in the centre of the detector and the \(z\)-axis along the beam pipe.
The \(x\)-axis points from the IP to the centre of the LHC ring,
and the \(y\)-axis points upwards.
Cylindrical coordinates \((r,\phi)\) are used in the transverse plane,
\(\phi\) being the azimuthal angle around the \(z\)-axis.
The pseudorapidity is defined in terms of the polar angle \(\theta\) as \(\eta = -\ln \tan(\theta/2)\).
Angular distance is measured in units of \(\Delta R \equiv \sqrt{(\Delta\eta)^{2} + (\Delta\phi)^{2}}\).} It consists of an inner tracking detector surrounded by a thin superconducting solenoid, electromagnetic (EM) and hadronic calorimeters, and a muon spectrometer. The inner tracking detector covers the pseudorapidity range \(|\eta| < 2.5\). It consists of silicon pixel, silicon microstrip, and transition radiation tracking (TRT) detectors. Lead/liquid-argon (LAr) sampling calorimeters provide electromagnetic energy measurements with high granularity.
A hadron calorimeter covers the central pseudorapidity range (\(|\eta| < 1.7\)).
The endcap and forward regions are instrumented with LAr calorimeters
for both the EM and hadronic energy measurements up to \(|\eta| = 4.9\).
The muon spectrometer surrounds the calorimeters and is based on
three large superconducting air-core toroidal magnets with eight coils each.
The field integral of the toroids ranges between \num{2.0} and \qty{6.0}{\tesla\metre}
across most of the detector.
The muon spectrometer includes a system of precision chambers for tracking and fast detectors for triggering.
 
A two-level trigger system~\cite{ATLAS:trigperf} was used during the data taking. The first-level trigger (L1) is implemented in hardware and uses a subset of the detector information. This is followed by a high-level software-based trigger that runs an algorithm similar to that in the offline reconstruction software, reducing the event rate from a maximum L1 rate of 100~kHz to approximately 1~kHz.
 
The trigger used in this search selects events with one or two electrons or muons, or a high-energy photon. During the highest instantaneous luminosity period, the minimum transverse momentum (\pt) threshold was 26~\GeV\ for the single-electron trigger and 17~\GeV\ for each electron in the dielectron trigger. The threshold for the single-muon trigger was also 26~\GeV, while asymmetric \pt thresholds of 22~\GeV\ and 8~\GeV\ were used for the dimuon trigger. The lowest transverse energy threshold for a single photon was 140~\GeV. Higher-threshold triggers with looser lepton identification criteria complement these lowest-threshold triggers. In the signal region chosen for the analysis, the trigger efficiencies range from 94\% to 100\% for simulated events with a spin-0 or spin-2 resonance having a mass between 220~\GeV\ and 3400~\GeV. After applying data quality requirements, the dataset of $\sqrt{s} = 13$~\TeV\ $pp$ collisions used in this search has an integrated luminosity of 140~\invfb~\cite{DAPR-2021-01}. The average number of $pp$ interactions per beam crossing (\pileup) ranged from ${\sim}13$ in 2015 to ${\sim}39$ in 2018. The peak instantaneous luminosity was \peakLumi.
 
An extensive software suite~\cite{ATL-SOFT-PUB-2021-001} is used in data simulation, in the reconstruction
and analysis of real and simulated data, in detector operations, and in the trigger and data acquisition
systems of the experiment.


\section{Event simulation}
\label{sec:MCsamples}

Samples of Monte Carlo (MC) simulated events are used to optimize the search strategy, evaluate the selection efficiency and study the different background contributions. The generated samples of signal events were processed with a detailed ATLAS detector simulation~\cite{SOFT-2010-01} based on \GEANT~\cite{Agostinelli:2002hh}. The \POWHEGBOX[v1]~\cite{Nason:2004rx,Frixione:2007vw, Alioli:2010xd,Bagnaschi:2011tu}  generator and \CT[10] PDF set~\cite{Lai:2010vv} were used to simulate gluon--gluon fusion (ggF) production of spin-0 resonances with masses of 200, 300, 500, 1000, 1500, 2000, 2500, 3000, and 3500~\GeV\  and an intrinsic width of 4~\MeV. This is much smaller than the experimental resolution (see Section~\ref{sec:signalBkgModel}) and is referred to as the narrow-width assumption.
Spin-2 resonances with masses of 200, 250, 300, 500, 750, 1000, 1500, 2000, 2500, 3000,  and 3500~\GeV\ and an intrinsic width of 4~\MeV\ were simulated for gluon--gluon and
quark--antiquark initial states.  These event samples were simulated at leading order (LO) in QCD in the Higgs Characterisation Model~\cite{Artoisenet:2013puc}
with \MGNLO[2.3.3]~\cite{Alwall:2014hca}.  For the high-mass spin-0 (spin-2) resonances, the parton showering, hadronization and multi-parton interactions were simulated with \PYTHIA[8.186]~\cite{Sjostrand:2007gs} using the AZNLO (A14) set of tuned parameters and the \CTEQ[6L1]~\cite{Pumplin:2002vw} (\NNPDF[2.3]~\cite{Ball:2012cx}) PDF set. Interference between the resonant signal and the non-resonant background is neglected because of the assumed narrow width of the resonance.

The SM $Z{\,+\,}\gamma$ process was simulated using the \Sherpa[2.2.2]~\cite{Gleisberg:2008ta} generator. The matrix elements were calculated using the \textsc{Comix}~\cite{Gleisberg:2008fv} and \OPENLOOPS[1.3.1]~\cite{Cascioli:2011va} generators. For real emission of up to three partons at LO in QCD, the matrix elements were merged with the \Sherpa parton shower~\cite{Schumann:2007mg} using the \MEPSatLO prescription~\cite{Hoeche:2009rj}. In addition, the \NNPDF[3.0nnlo] PDF set was used in conjunction with dedicated parton shower tuning developed by the \Sherpa authors. To study the background model in detail, a large sample of $Z{\,+\,}\gamma$ events was simulated  using fast simulation of the calorimeter response~\cite{fastsim}. The subdominant background process, $Z{\,}+$\,jets production, is modelled by using a control region enhanced in data events with jets misidentified as photons.


\section{Object and event selections}
\label{sec:eventSelection}

 
Events containing at least one primary vertex candidate formed by reconstructed tracks
with $\pt > 500$~\MeV\  are selected. The primary vertex candidate with the largest sum of the
squared transverse momenta ($\sum\pt^{2}$) of the associated tracks is considered to be the primary
vertex of the interaction of interest.  The $X\to Z(\to\ell\ell)\gamma$ candidate events are selected by requiring two same-flavour
opposite-charge leptons to form a $Z$ boson candidate, and at least one photon candidate.
 
Muon candidates are reconstructed by combining tracks in the inner tracking detector with tracks in
the muon spectrometer.  They are required to satisfy $\pt > 10$~\GeV\ and $|\eta| <2.7$. Muons must meet the \textit{Medium} identification criteria~\cite{MUON-2018-03}.
Electron and photon candidates are reconstructed from clusters of energy deposits in the EM calorimeter cells. Electron candidates must have a matching track reconstructed in the inner tracking detector and $\pt > 10$~\GeV, and be within the fiducial region of $|\eta|<2.47$, excluding the EM calorimeter barrel/endcap transition region of $1.37 < |\eta| < 1.52$~\cite{EGAM-2018-01}.
 
In the high-mass $X\to Z \gamma$ search, the energy deposits from the two electrons are very close together in the EM calorimeter, causing significant signal efficiency losses when using \textit{Loose} identification criteria~\cite{EGAM-2018-01}.
In addition, the sub-leading electron is often misreconstructed as a photon. This identification challenge is addressed by developing a dedicated MVA-based identification (MVA ID) criterion using a set of shower shape and track-based variables. The calorimeter shower shape variables used are $R_{\eta}$, $R_{\text{had}}$, $R_{\phi}$, $\Delta E_{s}$ and $E_{\textrm{ratio}}$. The track-based variables considered are $E/p$, eProbabilityHT, ${\Delta\eta_{1}}$, $d_{0}$, $n_{\text{innermost}}$, $n_{\text{Pixel}}$ and $n_{\text{Si}}$. The aforementioned variables are also used to develop standard electron identification criteria for the ATLAS Collaboration and are described in detail in Table~1 of Ref.~\cite{EGAM-2018-01}.
Additionally, $n_{\text{TRT}}$, defined as the number of hits in the TRT detector, and the  $\Delta\phi_{\text{rescaled}}$ variables, defined as the $\Delta\phi$ between the energy cluster and associated track in the presampler and the calorimeter's first, second and third layers after rescaling the EM energy deposits, are used in the MVA ID development.
All these input variables were chosen because of their separation power when comparing $m_{X}=5$~\TeV\ signal MC events to background sideband data excluding events with dielectron invariant mass within $\pm15$~\GeV\ of the $Z$ boson mass, $m_{Z}=91.2$~\GeV. The background in the sideband region includes both real and misreconstructed electrons. The MVA method is based on a Gradient Boosting Decision Tree (GBDT) architecture provided by the XGBoost software library~\cite{xGB} and the training is carried out on electrons in the signal and background events. Good  separation between signal and background is obtained with an optimal cut-point which corresponds to 99\% signal efficiency and 76\% background rejection. Among all training variables, the most discriminating shower shape and additional track-based variables are $E_{\text{ratio}}$ and $\Delta\phi_{\text{rescaled2}}$, respectively. The MVA ID is combined with the \textit{Loose} identification criterion by using a logical OR to cover the whole explored mass range, and this is named Mixed ID.\
Compared to the \textit{Loose} identification, the Mixed ID improves the identification efficiency of signal events by 6.2\% to 12.7\% across the full range of resonance masses. In order to form the $Z$ boson candidate, electrons misreconstructed as a photon candidate are also selected. Similarly to electron candidates, they are required to have $\pt > 50$~\GeV\ and be within $|\eta|<2.47$, excluding the region of $1.37 < |\eta| < 1.52$. They are also required to have at least one track with an angular distance $\Delta R<0.1$ from the photon.
 
A track-based isolation requirement~\cite{MUON-2018-03,EGAM-2018-01} is applied to the electrons and muons to make sure they are isolated from additional activity in the detector. Electron and muon candidates are also required to be associated with the primary vertex by requiring the longitudinal impact parameter, $ z_{0}$, to satisfy $| z_{0} \sin\theta|  < 0.5$~mm, where $\theta$ is the polar angle of the track.
The significance of the transverse impact parameter $\dzero$, calculated with respect to the measured beam-line position, must satisfy $|\dzero / \sigma_{\dzero}| < 3\,(5)$ for muons (electrons) where $\sigma_{\dzero}$ is the uncertainty in $\dzero$.
 
The $Z$ boson candidates are reconstructed from two same-flavour opposite-sign lepton candidates. For events where one of the electrons is misreconstructed as a photon, additional criteria are applied to increase the selection efficiency of the real electron and to reduce other photon backgrounds, which are determined from MC information. The angular distance between the selected electron and photon, $\Delta R$, must be less than 1.0. In addition, the relative difference between the transverse momenta of the two objects must be greater than 5\%, which suppresses the selection of objects that come from the same electron. If multiple $Z$ candidates are reconstructed in the same event, the one with mass closest to the $Z$ pole, $m_{Z}=91.2$~\GeV, is chosen. Corrections are applied to the four-momenta of the leptons to improve the resolution of signal events. In particular, muon momenta are corrected for collinear final-state radiation (FSR) effects~\cite{HIGG-2018-28}, and lepton (both electron and muon) four-momenta are corrected using a $Z$ boson mass constraint~\cite{HIGG-2013-21}. The $Z$ mass constraint improves the $\ell\ell\gamma$ mass resolution significantly. This is especially so in the muon channel, where the resolution improves by 8\% for the lowest resonance mass and by up to $70\%$ for larger masses, where the precision of the momentum measurement decreases with increasing muon transverse momentum. The corrected dilepton invariant mass is required to be within $\pm 15$~\GeV\ of the $Z$ boson mass.
 
Photon candidates are required to have $\pt > 15$~\GeV\ and pseudorapidity within the regions $|\eta|<1.37$ or $1.52<|\eta|<2.37$. The \textit{Tight} identification~\cite{EGAM-2018-01} and \textit{Loose} isolation criteria are also applied. The $Z(\ell\ell)\gamma$ candidate is reconstructed from the selected $Z$ boson and the photon with the largest transverse momentum. The selected lepton pair and photon are required to match the trigger object(s) used to select the event, with the lepton \pt requirement(s) being slightly higher than the trigger threshold(s).

To resolve potential ambiguities due to a single detector response being assigned to two objects by the reconstruction algorithm, an overlap removal procedure is applied. For electrons, if the leading object has $\pt <  500$~\GeV, other electron energy clusters closer than $|\Delta\eta|= 0.075$ and $|\Delta\phi| = 0.125$ are removed. If the leading electron has $\pt >  500$~\GeV, other electron energy clusters closer than $|\Delta\eta|= 0.05$ and $|\Delta\phi| = 0.05$ are removed. If an electron candidate's track is closer than $\Delta R = 0.02$ to a muon candidate's track, the electron is rejected. Photon candidates within $\Delta R =0.3$ of the selected lepton pair  are rejected;  this suppresses the FSR $Z{\,+\,}\gamma$ events and additional possible contributions from photons misidentified as electrons. If an electron is misreconstructed as a photon because of its closeness to another electron, no overlap removal is applied.

In order to further reduce the non-resonant $Z{\,+\,}\gamma$ background contamination, a requirement $\pt / m_{Z\gamma} > 0.2$ is placed on transverse momentum of the photon candidate relative to the invariant mass of the final-state particles, $m_{Z\gamma}$, where the latter must satisfy $200 < m_{Z\gamma} < 3500$~\GeV. The full analysis selections are summarized in Table~\ref{tab:evtSel}.
 
\begin{table}[htbp]
\caption{Object and event selections for the analysis.} 
\centering 
\small
\begin{tabular}{c c c c c} 
\hline\hline 
Selection                       & Muon                       & Electron                   & Electron as photon                                                                    & Photon               \\
\hline
$p_{\text{T}}$                  & $> 10$ GeV                   & $> 10$ GeV                   & $> 50$ GeV                                                                              & $> 15$ GeV             \\
\hline
\multirow{2}{*}{$|\eta|$}       & \multirow{2}{*}{$< 2.7$}     & $< 2.47$                     & $< 2.47$                                                                                & $< 2.37$               \\
&                            & Exclude [1.37, 1.52]       & Exclude [1.37, 1.52]                                                                  & Exclude [1.37, 1.52] \\
\hline
$|d_{0}|/\sigma_{d_{0}}$        & $< 3$                        & $< 5$                        &                                                                                       &                      \\
\hline
$|z_{0} \sin\theta|$     & $< 0.5$ mm                   & $< 0.5$ mm                   &                                                                                       &                      \\
\hline
Identification                              & \textit{Medium}            & Mixed                      & MVA                                                                                   & \textit{Tight}       \\
\hline
Isolation                       & Track-based \textit{Tight} & Track-based \textit{Tight} &                                                                                       & \textit{Loose}       \\
\hline
$\Delta R(\text{track},\gamma)$ &                            &                            & $< 0.1$                                                                                 &                      \\
\hline
$ee$ or $\mu\mu$ pair          & \multicolumn{2}{c}{$\ge 2$, opposite charge}  &                                                                                       &                      \\
\multirow{2}{*}{$e\gamma$ pair} &                            &                            & $\Delta R(e,\gamma) < 1 $                                                             &                      \\
&                            &                            & $|p_{\text{T}}^{e} - p_{\text{T}}^{\gamma}|/p_{\text{T}}^{e\text{ or }\gamma} > 5\% $ &                      \\
\hline
Categorization &\multicolumn{4}{c}{lepton pair closest to $m_{Z}=91.2$~GeV, decide if electron channel or muon channel}\\
\hline
\multirow{3}{*}{Event selections}& \multicolumn{4}{c}{$|m_{\ell\ell}^{\text{corrected}}-m_{Z}|<15$~GeV, $m_{Z}=91.2$~GeV} \\
& \multicolumn{4}{c}{Trigger match, overlap removal} \\
& \multicolumn{4}{c}{$p_{\text{T}}^{\gamma}/m_{Z\gamma}>0.2$;  signal region: $200 < m_{Z\gamma} < 3500$~GeV}\\
\hline
\hline
\end{tabular}
\label{tab:evtSel} 
\end{table}
 
\begin{figure}
\begin{center}
\includegraphics[width=0.6\textwidth]{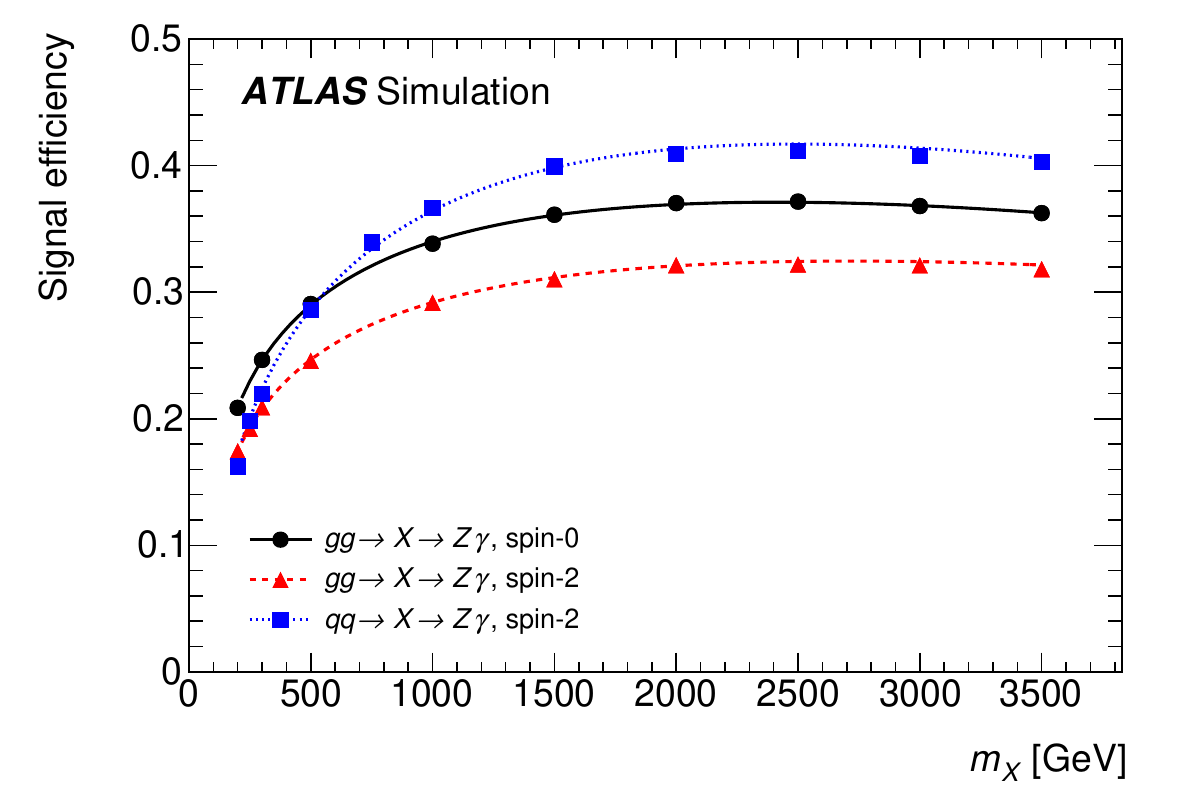}
\end{center}
\caption{Efficiency of $X\to Z\gamma$ final-state reconstruction and selection (including kinematic acceptance) as a function of resonance mass $m_X$ for spin-0 resonances generated via gluon--gluon fusion, and for spin-2 resonances generated from gluon--gluon and quark--antiquark initial states. The markers show the efficiencies for simulated events, while the curves indicate the parameterizations used in the analysis. The efficiencies are for $X\to Z\gamma$ where the $Z$ boson decays into $ee,\,\mu\mu\text{ or }\tau\tau$.}
\label{fig:sigEff}
\end{figure}

The signal efficiency is defined as the ratio of the number of events satisfying all selection criteria (as described above) to the total number of events expected from the process $pp\to X\to Z\gamma$, where the $Z$ boson decays into $ee,\,\mu\mu\text{ or }\tau\tau$. The efficiency is parameterized as a function of the resonance mass to interpolate efficiencies in the mass intervals between the simulated samples. This is done using the sum of a first-order polynomial and a logarithmic function, $\varepsilon=a+b\cdot m_{X}+c\cdot\ln(m_{X}+d)$, where $a$, $b$, $c$ and $d$ are free parameters in the fit. Figure~\ref{fig:sigEff} shows the reconstruction and selection efficiency for simulated spin-0 and spin-2 resonance events as a function of \mX. The efficiencies for spin-0 and spin-2 resonances produced via the gluon--gluon fusion process and spin-2 resonances originating from quark--antiquark initial states are parameterized separately. The efficiencies range from 22\% to 36\% over the mass range from 220~\GeV\ to 3400~\GeV\ for spin-0 resonances produced by gluon--gluon fusion. Differences in the production and decay of the spin-0 and spin-2 resonances result in significantly different transverse momentum and pseudorapidity distributions of the final-state $Z$ bosons and photons, leading to the differences between the detection efficiency curves shown in Figure~\ref{fig:sigEff}.

\section{Signal and background modelling}
\label{sec:signalBkgModel}

Analytic models are used to extract the signal and background yields from the $m_{Z\gamma}$ distribution of the data. Simulated signal samples are used to determine the parameters describing the signal shapes.
The models used to describe the background shapes are chosen by studying the simulated background samples, and the values of free parameters are obtained by fitting the models to the data.
 
\label{sec:signalModel}
 
The signal mass distribution of the $Z\gamma$ final states is well modelled by a double-sided Crystal Ball (DSCB) function (a Gaussian function with a power-law tail on each side)~\cite{dscb1, dscb2}. The Gaussian component of the signal distribution is described by the peak position $\mu_{\mathrm{CB}}$ and width $\sigma_{\mathrm{CB}}$.
 
\begin{figure}
\begin{center}
\includegraphics[width=.50\textwidth]{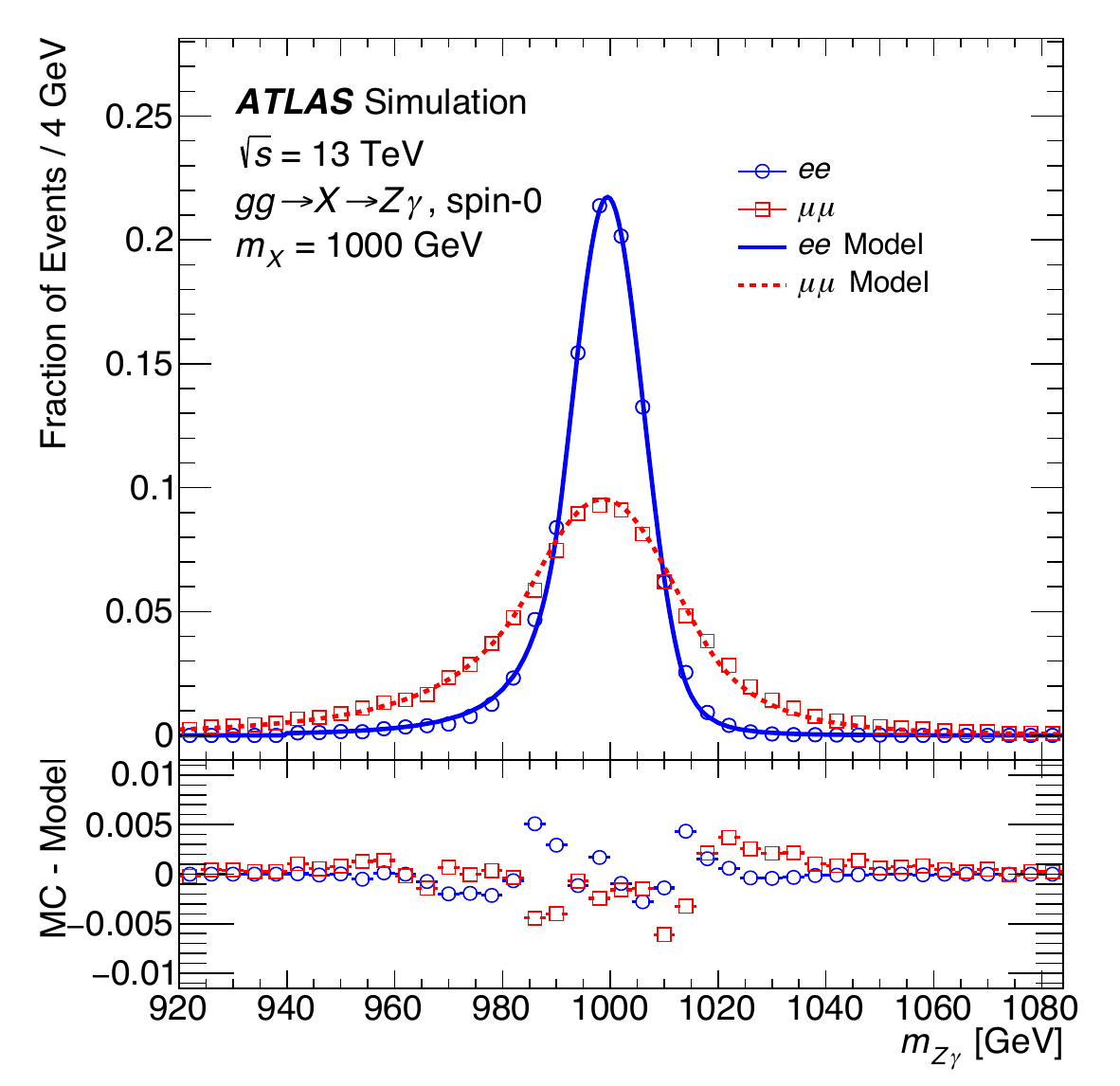}
\end{center}
\caption{Differential distributions of the invariant mass $m_{Z\gamma}$ of the spin-0 resonance with $m_X=1000$~\GeV\ produced by the gluon--gluon fusion process. The markers show the distributions of the simulated events. The solid and dashed lines are the fitted models in the $ee\gamma$ and $\mu\mu\gamma$ channels, respectively. The mass resolution in the muon channel is compatible to that in the electron channel when $m_X<300$ GeV.}
\label{fig:sig-mmg-fit}
\end{figure}

For the interpolation of the DSCB signal shape parameters as a function of \mX, all simulated signal events are fitted simultaneously to parameterize the signal shapes for mass points \mX\ between simulated samples. A set of polynomials whose coefficients are determined during the fitting process are used to interpolate signal shape parameters as a function of \mX. The parameterization is carried out separately for each of the three models considered, i.e.\ spin-0 and spin-2 resonances produced via gluon--gluon fusion and spin-2 resonances from quark--antiquark initial states, and for $Z\to ee$ and $Z\to\mu\mu$ final states. The $m_{Z\gamma}$ distributions of the spin-0 resonance at $\mX = 1000$~\GeV\ are shown in Figure~\ref{fig:sig-mmg-fit}. The simulated signal events and the fitted parametric models agree well, with differences below 5~per~mille. Good quality fits are also obtained for all other resonance models.
 
\label{sec:bckgModel}
 
The background consists mainly of non-resonant associated production of a $Z$ boson and a photon (irreducible background) and $Z$\,+\,jet events where the jet is misidentified as a photon (reducible background). Their relative contributions are determined by a simultaneous binned fit to the calorimeter isolation distribution of the photon candidate in the signal region and in a control region enriched in $Z$\,+\,jets background.
The control region is defined by requiring the photon candidate to fail the \textit{Tight} identification but pass a modified loose identification~\cite{EGAM-2018-01}. The calorimeter isolation distributions of the photon in the signal and control regions are determined by the simulated non-resonant $Z{\,+\,}\gamma$ samples, while the distributions of the misidentified jet are determined in the fit and assumed to be the same in the signal and control regions.
The composition is estimated separately in the $Z\to ee$ and $Z\to\mu\mu$ final states. In the electron (muon) channel, the ratio of $Z{\,+\,}\gamma$ events to all background events is 0.919 (0.908). Besides estimating the $Z{\,+\,}\gamma$ event fraction inclusively in the full mass region, it is also evaluated as a function of $m_{Z\gamma}$. The fractions are relatively stable, with the largest variation being 5\%.
Only simulated $Z{\,+\,}\gamma$ samples are used to construct the total background model to reduce the statistical fluctuations from the limited number of data-derived $Z$\,+\,jet events, which are obtained by requiring the photon candidate to satisfy the \textit{Loose}, but not the \textit{Tight}, identification criterion.
To account for the contribution from $Z$\,+\,jet events, the ratio of $Z$\,+\,jet to $Z{\,+\,}\gamma$ events is fitted as an exponential function of $m_{Z\gamma}$. The functional form is chosen to have the minimum spurious signal described below. The total background is obtained after multiplying the $Z{\,+\,}\gamma$ distribution in MC simulation by this exponential function to estimate the $Z$\,+\,jet contribution. The fit uncertainty of the ratio is propagated to the total background as well. The background distribution falls smoothly as a function of $m_{Z\gamma}$.

The analytic background models are chosen so as to reduce the bias in the extracted signal yield and also by limiting the number of free parameters in the fits to avoid a reduction in sensitivity~\cite{HIGG-2012-27}. For each analysis category, the bias (also known as the \enquote{spurious signal}) is estimated as a function of $m_{X}$ by fitting the $m_{Z\gamma}$ distribution of the background, obtained as described above, with signal-plus-background models. The spurious signal is required to be less than 50\% of the expected statistical uncertainty of the signal yield. The function with the fewest free parameters is selected if this requirement is satisfied by two or more of the functions considered. The Dijet function\footnote{The Dijet function is defined as $f_\mathrm{bkg}(x;b,a_{0}) = N (1-x)^{b}x^{a_{0}}$, where $x= m_{Z\gamma}/\sqrt{s}$, $N$ is the normalization factor, and $b$ and $a_{0}$ are free parameters in each fit.} is selected for both the electron and muon channels. The envelope of the spurious signal is used to define a systematic uncertainty of the background modelling, parameterized as an exponential function of $m_{Z\gamma}$. For the $Z\to ee$ ($Z\to\mu\mu$) final state, the spurious signal ranges from 10.2 (10.7) events at 220~\GeV to 0.003 (0.007) events at 3400~\GeV. Signal models for different-spin resonances are tested in the fits, and the parameterized spurious-signal uncertainties derived with spin-0 resonance samples are the most conservative. If the purity of the $Z{\,+\,}\gamma$ sample is varied by $\pm 5\%$, or the fitted ratio of $Z$\,+\,jet to $Z{\,+\,}\gamma$ events as a function of $m_{Z\gamma}$  is varied by $\pm1\sigma$ of the error, the Dijet function is still selected for both channels. This indicates that the search relies on the parameterization of the $m_{Z\gamma}$ distribution, but only mildly on the background composition.


\section{Systematic uncertainties}
\label{sec:sys}

 
The dominant systematic uncertainty is the spurious signal defined as the bias induced in the signal yield by the choice of a particular background model. Its evaluation is described in Section~\ref{sec:signalBkgModel}, where it is found be as large as 10.2~events in the electron channel and 10.7~events in the muon channel.
 
The uncertainty in the combined 2015--2018 integrated luminosity is 0.83\%~\cite{DAPR-2021-01}, obtained using the LUCID-2 detector~\cite{Avoni:2018iuv} for the primary luminosity measurements, complemented by measurements using the inner detector and calorimeters. The systematic uncertainties impacting the signal modelling come from the muon momentum scale and resolution, and the electron and photon energy scales and resolutions. Their impact on the peak position ($\mu_{\text{CB}}$) and width ($\sigma_{\text{CB}}$) of the simulated signal distribution is estimated from the relative changes in the fitted $m_{Z\gamma}$ signal distribution when varying the momentum or energy scales and resolutions by their uncertainties. The muon momentum scale and resolution systematic uncertainties are determined from $Z\to\mu\mu$ and $J/\psi\to\mu\mu$ events using the techniques described in Ref.~\cite{Aad:2016jkr}. The muon momentum scale uncertainty and the sagitta bias~\cite{MUON-2022-01} lead to a $\mu_\mathrm{CB}$ uncertainty of up to 0.023\%, while both the muon spectrometer and muon identification contribute to the muon momentum resolution and lead to uncertainties in $\sigma_\mathrm{CB}$ of up to 1.9\% and 1.8\% respectively. The systematic uncertainties in the electron and photon energy scale and resolution follow those in Refs.~\cite{Aad:2014nim, ATLAS-egammacalib}. The overall energy scale factors and their uncertainties were determined using $Z\to ee$ events collected during 2015 and 2016. Compared to Ref.~\cite{ATLAS-egammacalib}, several systematic uncertainties were re-evaluated with the 13~\TeV\ data, including uncertainties related to the observed LAr cell non-linearity, the detector material simulation, the intercalibration of the first and second layers of the calorimeter, and the pedestal corrections.
The electron/photon energy resolution uncertainties produce an uncertainty in $\sigma_\mathrm{CB}$ varying from 2.5\% to 10\% in the muon channel and from 7\% to 60\% in the electron channel. The variation in $\mu_\mathrm{CB}$ due to the electron and photon energy scale systematic uncertainty is less than 0.4\% for the muon channel and less than 0.7\% for the electron channel. The systematic uncertainties in the signal efficiency due to the reconstruction, identification, isolation and trigger efficiencies for leptons and photons are estimated in simulation from the relative change in the signal efficiency when each of those efficiencies is varied by its uncertainty. The photon triggers, identification and isolation contribute a total systematic uncertainty of up to 1.5\% (1.7\%) to the signal efficiency in the muon (electron) channel. The electron reconstruction, identification, isolation and trigger contribute a systematic uncertainty of up to 4\% to the signal efficiency in the electron channel. In the muon channel, the signal efficiency systematic uncertainties from the muon triggers, reconstruction and isolation are estimated to not exceed 1\%, 6\% and 1.2\%, respectively. All these systematic uncertainties affecting the signal efficiency are estimated using spin-0 resonance samples only and are also used in the spin-2 resonance results. To check whether a bias could be introduced by this uncertainty assignment, the largest systematic uncertainty in the signal efficiency (i.e.\ the muon reconstruction efficiency) is also estimated using spin-2 resonance samples. The estimates of this systematic uncertainty from the spin-0 and spin-2 resonance samples are compatible within the statistical uncertainty.
An \enquote{extra smearing} muon \pt uncertainty accounts for the impact of the poorly measured resolution of high-$p_{\text{T}}$ muons (usually $>300$~\GeV) which satisfy the \textit{Medium}, but not the
\textit{HighPt}~\cite{MUON-2018-03}, identification criterion. The impact on the $m_{\mu\mu\gamma}$ resolution is estimated to be 2.4\%.
The systematic uncertainty due to electron charge misidentification is evaluated using $Z\rightarrow ee$ events and found to be negligible.
 
The uncertainty due to the MVA and Mixed electron identifications applied to the reconstructed electrons in this high-mass analysis is evaluated by applying the tag-and-probe method to the $Z\rightarrow ee$ data and MC samples. Events where the electrons have an angular separation $\Delta R (e,e) < 0.5$ are used to imitate the case where the electrons are very close to each other, as in the high-mass resonance analysis. Due to the $\Delta R (e,e)$ requirement, the $Z\rightarrow ee$ event yield is limited and thus an inclusive \pt bin is defined in the EM calorimeter barrel and endcap region. The efficiencies are evaluated for electrons with $\Delta R (e,e) <  0.5$ in an inclusive \pt region between 27 and 3000~\GeV\ and the barrel ({|$\eta$}| < 1.37) and endcap ({1.52 < |$\eta$}| < 2.47) regions. The systematic uncertainty due to the MVA ID and its mixture with the \textit{Loose} ID is estimated to vary between 1.0\% and 1.1\% for resonance masses between 220~\GeV\ and 3400~\GeV.

Table \ref{sys_table} summarizes the estimated systematic uncertainties for the $X\to Z \gamma$ search in the $ee$ and $\mu\mu$ channels and the mass range $\mX = 220{-}3400~\GeV$.

\begin{table}
\caption{The main sources of systematic uncertainty for the $X\to Z \gamma$ search. The gluon--gluon fusion spin-0 signal samples
produced for $m_{X} = 220{-}3400$~\GeV\ are used to evaluate the systematic uncertainty.
The uncertainty ranges span the variations among different categories and different $m_{X}$ resonance masses. The uncertainty due to the spurious signal is reported as an absolute number of events. In the table, \enquote{ID} for photons and electrons refers to identification efficiency uncertainties, \enquote{ISO} refers to isolation efficiency uncertainties, \enquote{TRIG} refers to trigger efficiency uncertainties, \enquote{RECO} refers to muon reconstruction efficiency uncertainty and \enquote{TTVA} refers to muon track-to-vertex-association efficiency uncertainty.}
\centering
\begin{tabular}{ccc}
\hline\hline
Category& $\mu\mu\gamma$ & $ee\gamma$ \\
\hline\hline
Luminosity  & \multicolumn{2}{c}{0.83\%} \\
\hline
\multicolumn{3}{c}{\textit{Signal Efficiency}}  \\
\hline
Photon ID/ISO/TRIG efficiency & 1.0\%\,--\,1.5\%& 1.0\%\,--\,1.7\% \\
 
Muon ISO efficiency & 1.0\%\,--\,1.2\%&--\\
Muon RECO efficiency & 0.22\%\,--\,6\%\phantom{.00}&-- \\
Muon TTVA efficiency &0.14\%\,--\,0.23\%&-- \\
Muon TRIG efficiency & 0.6\%\,--\,1.0\%&-- \\
 
Electron ID/ISO/RECO/TRIG efficiency &-- &2.9\%\,--\,4\%\phantom{0.} \\
MVA/Mixed electron ID efficiency & -- & 1.0\%\,--\,1.1\%\\
Pile-up& $< 0.016\%$& --\\
\hline
\multicolumn{3}{c}{\textit{Signal modelling effect on} $\mu_\text{CB}$}  \\
\hline
Electron and photon energy scale  & 0.33\%\,--\,0.4\%\phantom{0} & 0.15\%\,--\,0.7\%\phantom{0}\\
Muon momentum scale/sagitta bias & $< 0.023\%$ &--\\
\hline
\multicolumn{3}{c}{\textit{Signal modelling effect on} $\sigma_\text{CB}$}  \\
\hline
Electron and photon energy resolution &2.5\%\,--\,10\%\phantom{.}&\phantom{0}7\%\,--\,60\%\\
Muon ID resolution		&0.4\%\,--\,1.8\%&--\\
Muon MS resolution 	 &0.6\%\,--\,1.9\%&--\\
Extra smearing of muon \pt 	 &2.4\%&--\\
\hline
\multicolumn{3}{c}{\textit{Background modelling (in number of events)}}  \\
\hline
Spurious signal &0.01\,--\,10.7&0.003\,--\,10.2\\
\hline\hline
\end{tabular}
\label{sys_table}
\end{table}


\section{Results}
\label{sec:result}

\begin{figure}[ht]
\begin{center}
\subfigure[]{\includegraphics[width=0.49\textwidth]{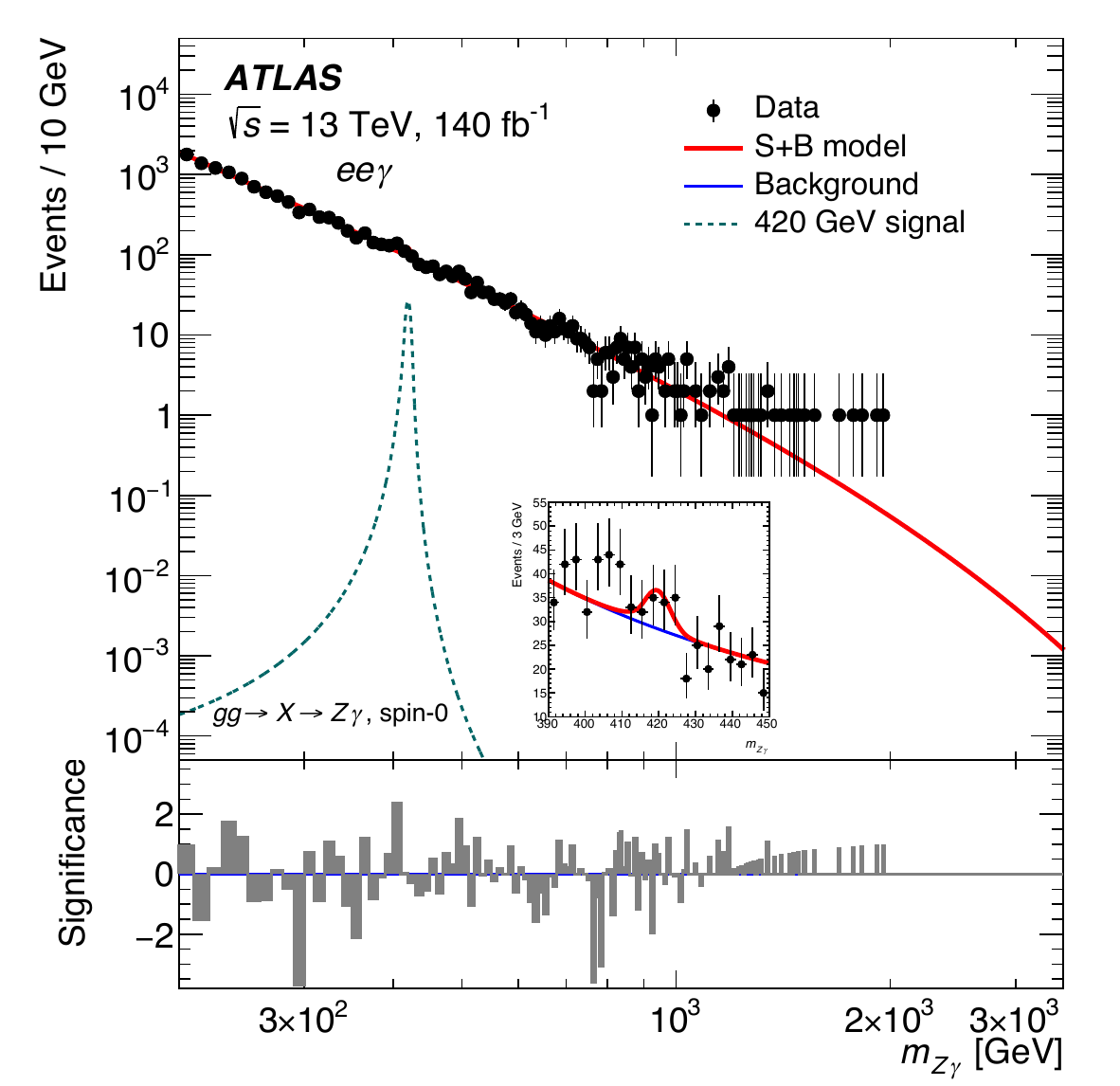}}
\subfigure[]{\includegraphics[width=0.49\textwidth]{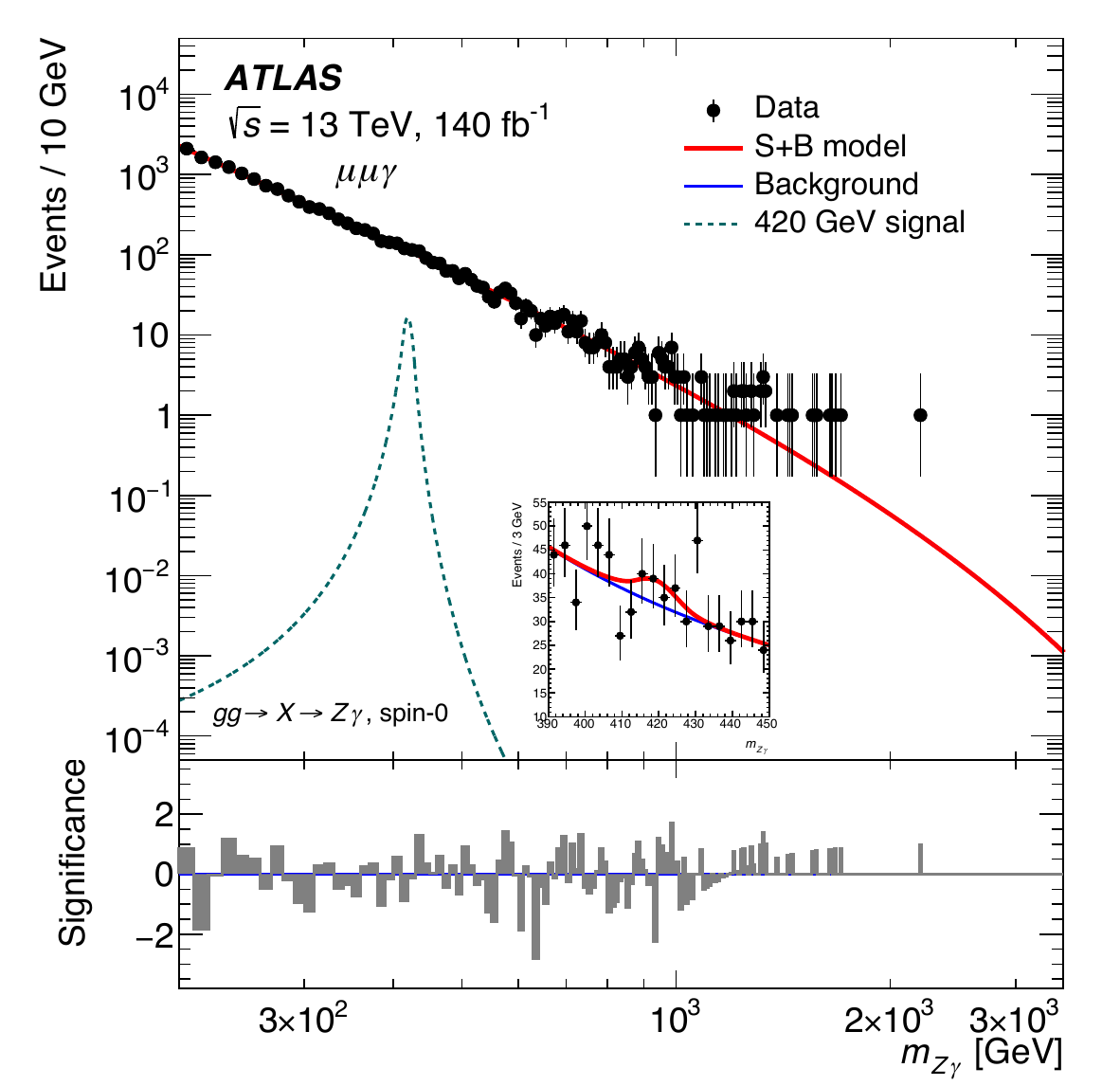}}
\end{center}
\caption{The $Z\gamma$ invariant mass distributions of data events satisfying the high-mass selection for the (a) $ee\gamma$ and (b) $\mu\mu\gamma$ channels. The points with error bars represent the data and statistical uncertainty. The background component (solid blue line in the inset) and spin-0 signal component (dashed dark cyan line) of the signal + background unbinned fit (solid red line) to data are displayed. The bottom panel of each figure shows the significance, which is defined as the residual of the data with respect to the fitted background component divided by the statistical uncertainty of the data. The lower mass region is expanded and displayed in the two inset plots, where an excess with a combined local significance of $2.3\sigma$ at 420~\GeV\ can be seen.}
\label{fig:bkgOnlyFit}
\end{figure}

An unbinned profile-likelihood-ratio fit method~\cite{Cowan:2010js} is used to estimate the heavy-resonance production cross section times the branching ratio of the $X\to Z\gamma$ decay, \sigxbr, in the mass range between 220~\GeV\ and 3400~\GeV.
In the likelihood function, the expected number of signal events $N_{\text{sig}}$ is defined as $N_{\text{sig}}=L\times\sigma(pp\to X)\times\mathcal{B}\bigl(X\to Z(\to ee, \mu\mu\text{ or }\tau\tau)\gamma\bigr)\times\varepsilon$, where $L$ is the integrated luminosity and $\varepsilon$ is the parameterized signal efficiency as a function of $m_{X}$. The invariant mass ($m_{Z\gamma}$) distributions of data events in both the $ee\gamma$ and $\mu\mu\gamma$ channels are fitted simultaneously with the signal-plus-background models and are shown in Figure~\ref{fig:bkgOnlyFit}. The highest-mass $ee\gamma$ and $\mu\mu\gamma$ events in the data are at 2.0~\TeV\ and 2.2~\TeV, respectively. No significant excess relative to the background-only hypothesis is seen.
For spin-0 heavy resonances, the largest excess is observed at 420~\GeV\ with a local significance of 2.3~standard deviations after combining the $ee\gamma$ and $\mu\mu\gamma$ channel distributions shown in Figure~\ref{fig:bkgOnlyFit}. The individual significances at the same $m_{X}$ value are 2.1$\sigma$ and 1.1$\sigma$ in the electron and muon channels respectively.

\begin{figure}[!htbp]
\begin{center}
\includegraphics[width=0.7\linewidth]{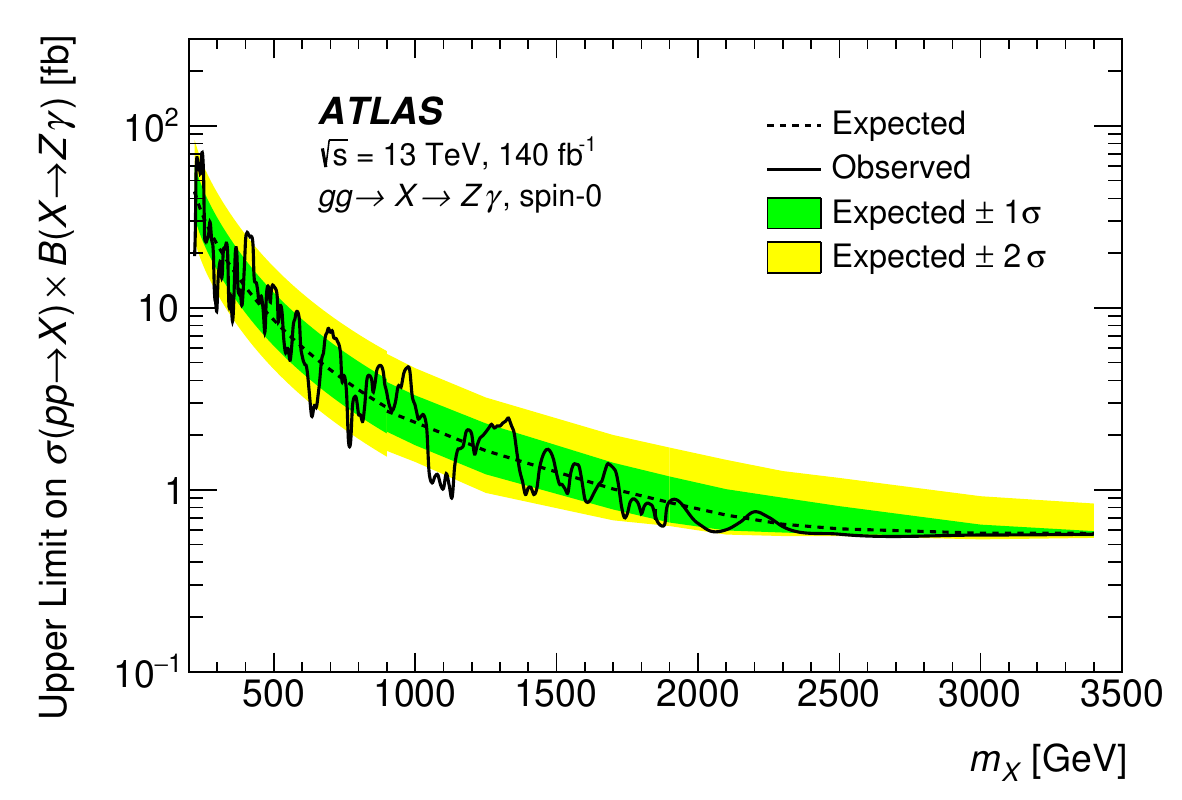}
\end{center}
\caption{Observed (solid line) and expected (dashed line) 95\% CL upper limits on the production cross section times branching ratio of a narrow-width spin-0 resonance $X$ produced from gluon--gluon initial states and decaying into a $Z$ boson and a photon, $\sigma(pp\to X)\cdot\mathcal{B}(X\to Z\gamma)$, as a function of the resonance mass $m_{X}$. Observed (expected) results are derived from ensemble tests for $m_{X}>1850\,(900)$~\GeV\ and from asymptotic formulae for lower $m_{X}$ values. The green and yellow bands correspond to the $\pm1\sigma$ and $\pm2\sigma$ intervals for the expected upper limits. The limits are shown in the $m_{X}$ range from 220~\GeV\ to 3400~\GeV\ and are obtained from the combined $ee\gamma$ and $\mu\mu\gamma$ channels. }
\label{fig:statistical_limits_ggH_spin0}
\end{figure}

The probability of compatibility between the data and the expected background plus signal is examined for increasing values of the signal cross section, and a modified frequentist (CL$_\text{s}$) approach~\cite{Read:2002hq} is used to set an upper limit on the cross section.
The limit at 95\% confidence level (CL) is determined by identifying the signal cross section for which the CL$_\text{s}$ value is equal to 0.05. The observed (expected) cross-section limits for $m_{X}$ up to 1850 (900)~\GeV\ are derived using closed-form asymptotic formulae~\cite{Cowan:2010js}.  At higher $m_{X}$ values the asymptotic formulae underestimate the observed (expected) limits by 5\% to 17\% (1.4\% to 29\%) because of the smaller number of events, and ensemble tests with sampling distributions generated using pseudo-experiments are used instead. Figure~\ref{fig:statistical_limits_ggH_spin0} shows the observed and expected upper limits as a function of $m_{X}$ for a spin-0 resonance signal produced via gluon--gluon fusion, using the combined data from the $ee\gamma$ and $\mu\mu\gamma$ channels. The observed (expected) limits range from 65.5~fb to 0.6~fb (43.3~fb to~0.6 fb).
The search is limited by the statistical uncertainty of the selected data events in the $m_{Z\gamma}$ distribution. The dominant systematic uncertainty is the spurious-signal uncertainty, which has at most a 12\% impact on the asymptotic expected upper limit on \sigxbr.
 
\begin{figure}
\begin{center}
\subfigure[]{\includegraphics[width=0.49\textwidth]{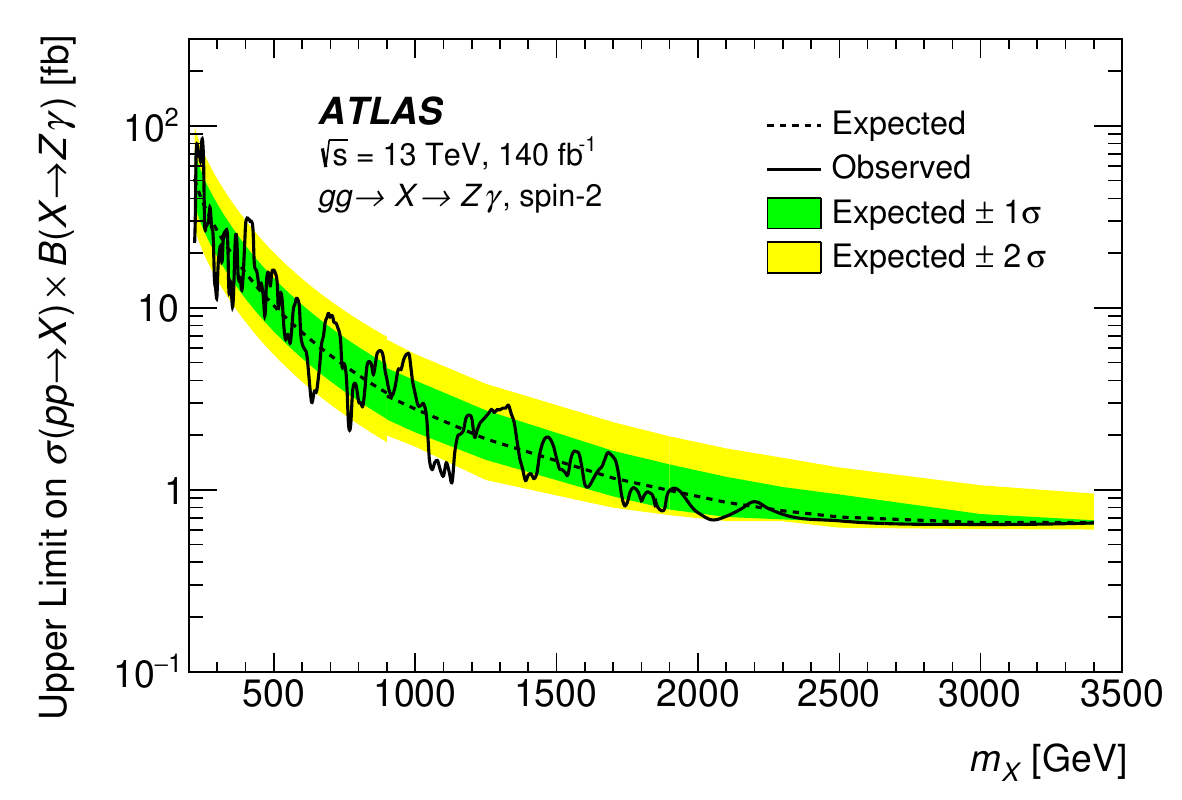}}
\subfigure[]{\includegraphics[width=0.49\textwidth]{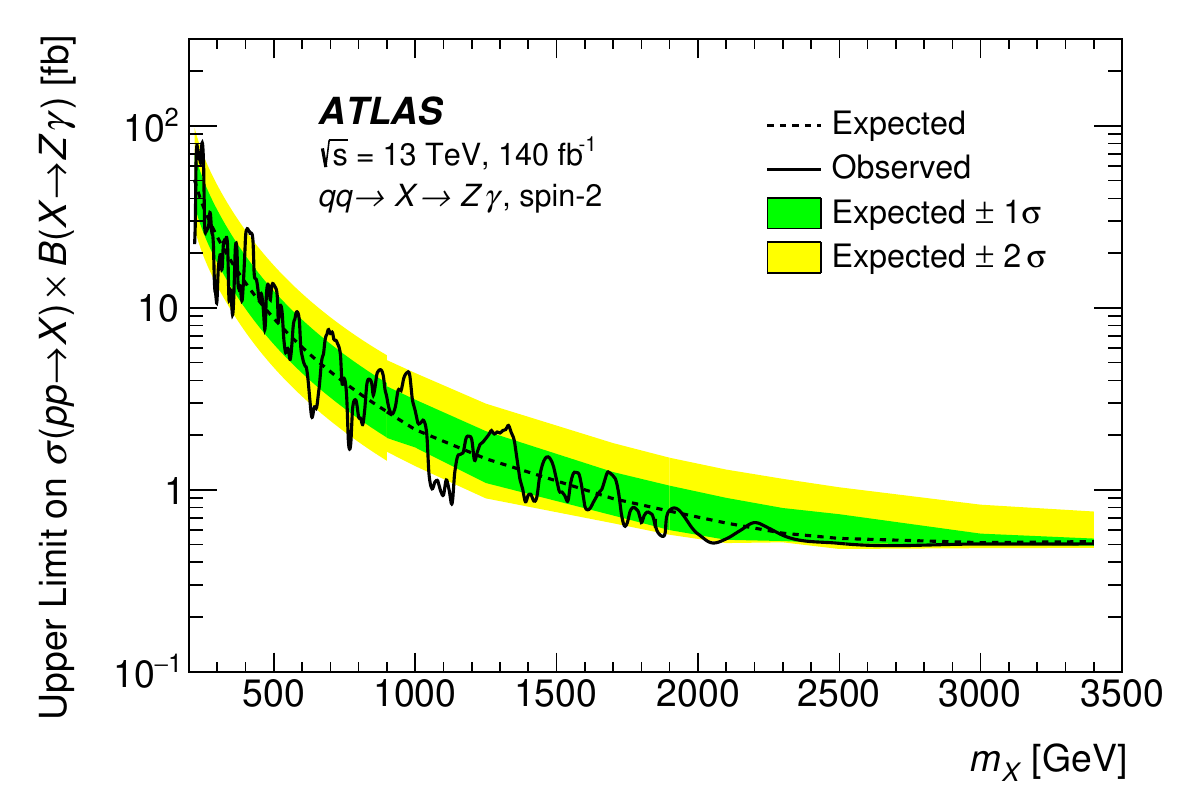}}
\end{center}
\caption{Observed (solid line) and expected (dashed line) 95\% CL limits on the production cross section times branching ratio of a narrow-width spin-2 resonance $X$ produced from (a) gluon--gluon initial states and (b) $q\bar{q}$ initial states and decaying into a $Z$ boson and a photon, $\sigma(pp\to X)\cdot\mathcal{B}(X\to Z\gamma)$, as a function of the resonance mass $m_{X}$. Observed (expected) results are derived from ensemble tests for $m_{X}>1850\,(900)$~\GeV\ and from asymptotic formulae for lower $m_{X}$ values. The green and yellow bands correspond to the $\pm1\sigma$ and $\pm2\sigma$ intervals for the expected upper limit respectively. The limits are shown in the $m_{X}$ range from 220~\GeV\ to 3400~\GeV\ and are obtained from the combined $ee\gamma$ and $\mu\mu\gamma$ channels.}
\label{fig:limits_spin2}
\end{figure}

The results are also interpreted in terms of spin-2 resonances (for both the $ggX$ and $q\bar{q}X$ processes) in the same mass range as the nominal spin-0 resonances. As shown in Figure~\ref{fig:limits_spin2}, the observed (expected) limits range from 77.4~fb to 0.6~fb (50.8~fb to 0.6~fb) for a $ggX$ spin-2 resonance and from 76.1~fb to 0.5~fb (50.3~fb to 0.5~fb) for a $q\bar{q}X$ spin-2 resonance. Table~\ref{tab:limit} summarizes the observed (expected) upper limits on \sigxbr\ for spin-0 and spin-2 heavy-resonance masses from 220~\GeV\ to 3400~\GeV.
 
\begin{table}
\caption{The observed (expected) upper limits on $\sigma(pp\to X)\cdot\mathcal{B}(X\to Z\gamma)$ for spin-0 and spin-2 heavy resonances at 95\% CL. The value of $m_{X}$ varies from 220~\GeV\ to 3400~\GeV.}
\centering
\begin{tabular}{ccc}
\hline\hline
95\% CL upper limits on \sigxbr\ & Observed & Expected \\
\hline
$ggX$ spin-0  & 65.5 fb -- 0.6 fb & 43.3 fb -- 0.6 fb\\
$ggX$ spin-2 & 77.4 fb -- 0.6 fb & 50.8 fb -- 0.6 fb\\
$q\bar{q}X$ spin-2 & 76.1 fb -- 0.5 fb & 50.3 fb -- 0.5 fb\\
\hline\hline
\end{tabular}
\label{tab:limit}
\end{table}


\FloatBarrier
 
\section{Conclusion}
\label{sec:conclusion}

A search for new resonances decaying into the $Z(\to\ell\ell)\gamma$ final state in the mass range between 220~\GeV\ and 3400~\GeV\ has been performed using 140~fb$^{-1}$ of $\sqrt{s}=13$~\TeV\ $pp$ collision data recorded with the ATLAS detector at the LHC.  The observed data are in agreement with the smoothly falling background predicted by the SM. No evidence of \xtoZg\ decay is observed, and upper limits are set on \sigxbr\ as a function of $m_X$. The results are presented using spin-0 and spin-2 interpretations. For spin-0 resonances, the observed limits vary between 65.5~fb and 0.6~fb. The cross-section limits vary between 77.4 (76.1)~fb and 0.6 (0.5)~fb for spin-2 resonances produced from gluon--gluon (quark--antiquark) initial states. 
These results improve the expected upper limit on \sigxbr\ for a spin-0 resonance by a factor of 1.9 to 4 in the $m_{X}$ range of 250~\GeV\ to 2400~\GeV\ covered by a previous ATLAS search.
In addition, this search extends the covered mass range to 3400~\GeV\ by using the higher integrated luminosity of the full \RunTwo dataset as well as an MVA electron identification technique. Compared to a resonance search by ATLAS using hadronic decays of the $Z$ boson and the full \RunTwo dataset, this search probes lower $m_{X}$ values, down to 220~\GeV, and has better sensitivity up to 2300~\GeV.


\section*{Acknowledgements}


We thank CERN for the very successful operation of the LHC, as well as the
support staff from our institutions without whom ATLAS could not be
operated efficiently.
 
We acknowledge the support of
ANPCyT, Argentina;
YerPhI, Armenia;
ARC, Australia;
BMWFW and FWF, Austria;
ANAS, Azerbaijan;
CNPq and FAPESP, Brazil;
NSERC, NRC and CFI, Canada;
CERN;
ANID, Chile;
CAS, MOST and NSFC, China;
Minciencias, Colombia;
MEYS CR, Czech Republic;
DNRF and DNSRC, Denmark;
IN2P3-CNRS and CEA-DRF/IRFU, France;
SRNSFG, Georgia;
BMBF, HGF and MPG, Germany;
GSRI, Greece;
RGC and Hong Kong SAR, China;
ISF and Benoziyo Center, Israel;
INFN, Italy;
MEXT and JSPS, Japan;
CNRST, Morocco;
NWO, Netherlands;
RCN, Norway;
MEiN, Poland;
FCT, Portugal;
MNE/IFA, Romania;
MESTD, Serbia;
MSSR, Slovakia;
ARRS and MIZ\v{S}, Slovenia;
DSI/NRF, South Africa;
MICINN, Spain;
SRC and Wallenberg Foundation, Sweden;
SERI, SNSF and Cantons of Bern and Geneva, Switzerland;
MOST, Taiwan;
TENMAK, T\"urkiye;
STFC, United Kingdom;
DOE and NSF, United States of America.
In addition, individual groups and members have received support from
BCKDF, CANARIE, Compute Canada and CRC, Canada;
PRIMUS 21/SCI/017 and UNCE SCI/013, Czech Republic;
COST, ERC, ERDF, Horizon 2020, ICSC-NextGenerationEU and Marie Sk{\l}odowska-Curie Actions, European Union;
Investissements d'Avenir Labex, Investissements d'Avenir Idex and ANR, France;
DFG and AvH Foundation, Germany;
Herakleitos, Thales and Aristeia programmes co-financed by EU-ESF and the Greek NSRF, Greece;
BSF-NSF and MINERVA, Israel;
Norwegian Financial Mechanism 2014-2021, Norway;
NCN and NAWA, Poland;
La Caixa Banking Foundation, CERCA Programme Generalitat de Catalunya and PROMETEO and GenT Programmes Generalitat Valenciana, Spain;
G\"{o}ran Gustafssons Stiftelse, Sweden;
The Royal Society and Leverhulme Trust, United Kingdom.
 
The crucial computing support from all WLCG partners is acknowledged gratefully, in particular from CERN, the ATLAS Tier-1 facilities at TRIUMF (Canada), NDGF (Denmark, Norway, Sweden), CC-IN2P3 (France), KIT/GridKA (Germany), INFN-CNAF (Italy), NL-T1 (Netherlands), PIC (Spain), ASGC (Taiwan), RAL (UK) and BNL (USA), the Tier-2 facilities worldwide and large non-WLCG resource providers. Major contributors of computing resources are listed in Ref.~\cite{ATL-SOFT-PUB-2023-001}.


\clearpage
 
\printbibliography
 
\clearpage
 
\begin{flushleft}
\hypersetup{urlcolor=black}
{\Large The ATLAS Collaboration}

\bigskip

\AtlasOrcid[0000-0002-6665-4934]{G.~Aad}$^\textrm{\scriptsize 102}$,
\AtlasOrcid[0000-0002-5888-2734]{B.~Abbott}$^\textrm{\scriptsize 120}$,
\AtlasOrcid[0000-0002-1002-1652]{K.~Abeling}$^\textrm{\scriptsize 55}$,
\AtlasOrcid[0000-0001-5763-2760]{N.J.~Abicht}$^\textrm{\scriptsize 49}$,
\AtlasOrcid[0000-0002-8496-9294]{S.H.~Abidi}$^\textrm{\scriptsize 29}$,
\AtlasOrcid[0000-0002-9987-2292]{A.~Aboulhorma}$^\textrm{\scriptsize 35e}$,
\AtlasOrcid[0000-0001-5329-6640]{H.~Abramowicz}$^\textrm{\scriptsize 151}$,
\AtlasOrcid[0000-0002-1599-2896]{H.~Abreu}$^\textrm{\scriptsize 150}$,
\AtlasOrcid[0000-0003-0403-3697]{Y.~Abulaiti}$^\textrm{\scriptsize 117}$,
\AtlasOrcid[0000-0002-8588-9157]{B.S.~Acharya}$^\textrm{\scriptsize 69a,69b,m}$
% ...
\bigskip
\\

% ...
$^{29}$Physics Department, Brookhaven National Laboratory, Upton NY; United States of America.\\
% ...
$^{35}$$^{(a)}$Facult\'e des Sciences Ain Chock, R\'eseau Universitaire de Physique des Hautes Energies - Universit\'e Hassan II, Casablanca;$^{(b)}$Facult\'{e} des Sciences, Universit\'{e} Ibn-Tofail, K\'{e}nitra;$^{(c)}$Facult\'e des Sciences Semlalia, Universit\'e Cadi Ayyad, LPHEA-Marrakech;$^{(d)}$LPMR, Facult\'e des Sciences, Universit\'e Mohamed Premier, Oujda;$^{(e)}$Facult\'e des sciences, Universit\'e Mohammed V, Rabat;$^{(f)}$Institute of Applied Physics, Mohammed VI Polytechnic University, Ben Guerir; Morocco.\\
% ...
$^{49}$Fakult\"{a}t Physik , Technische Universit{\"a}t Dortmund, Dortmund; Germany.\\
% ...
$^{55}$II. Physikalisches Institut, Georg-August-Universit\"{a}t G\"ottingen, G\"ottingen; Germany.\\
% ...
$^{69}$$^{(a)}$INFN Gruppo Collegato di Udine, Sezione di Trieste, Udine;$^{(b)}$ICTP, Trieste;$^{(c)}$Dipartimento Politecnico di Ingegneria e Architettura, Universit\`a di Udine, Udine; Italy.\\
% ...
$^{102}$CPPM, Aix-Marseille Universit\'e, CNRS/IN2P3, Marseille; France.\\
% ...
$^{117}$Department of Physics, New York University, New York NY; United States of America.\\
% ...
$^{120}$Homer L. Dodge Department of Physics and Astronomy, University of Oklahoma, Norman OK; United States of America.\\
% ...
$^{150}$Department of Physics, Technion, Israel Institute of Technology, Haifa; Israel.\\
$^{151}$Raymond and Beverly Sackler School of Physics and Astronomy, Tel Aviv University, Tel Aviv; Israel.\\
% ...
$^{a}$ Also Affiliated with an institute covered by a cooperation agreement with CERN.\\
$^{b}$ Also at An-Najah National University, Nablus; Palestine.\\
$^{c}$ Also at Borough of Manhattan Community College, City University of New York, New York NY; United States of America.\\
$^{d}$ Also at Center for High Energy Physics, Peking University; China.\\
$^{e}$ Also at Center for Interdisciplinary Research and Innovation (CIRI-AUTH), Thessaloniki; Greece.\\
$^{f}$ Also at Centro Studi e Ricerche Enrico Fermi; Italy.\\
% ...
$^{m}$ Also at Department of Physics, King's College London, London; United Kingdom.\\
% ...

\end{flushleft}


\end{document}